\documentclass[a4paper]{article}

\usepackage{INTERSPEECH2022}

\usepackage{hyperref}
\hypersetup{
    colorlinks=true,
    linkcolor=blue,
    filecolor=magenta,
    urlcolor=blue,
    citecolor=blue,
}

\let\OLDthebibliography\thebibliography
\renewcommand\thebibliography[1]{
  \OLDthebibliography{#1}
  \setlength{\parskip}{0pt}
  \setlength{\itemsep}{1.3pt plus 0.6ex}
}

\title{Vocal effort modeling in neural TTS for improving the intelligibility of synthetic speech in noise}
\name{Tuomo Raitio, Petko Petkov, Jiangchuan Li, Muhammed Shifas, Andrea Davis, Yannis Stylianou}

\address{Apple}

\email{\vspace{-5mm}}

\begin{document}

\maketitle
\begin{abstract}
We present a neural text-to-speech (TTS) method that models natural vocal effort variation to improve the intelligibility of synthetic speech in the presence of noise. The method consists of first measuring the spectral tilt of unlabeled conventional speech data, and then conditioning a neural TTS model with normalized spectral tilt among other prosodic factors. Changing the spectral tilt parameter and keeping other prosodic factors unchanged enables effective vocal effort control at synthesis time independent of other prosodic factors. By extrapolation of the spectral tilt values beyond what has been seen in the original data, we can generate speech with high vocal effort levels, thus improving the intelligibility of speech in the presence of masking noise. We evaluate the intelligibility and quality of normal speech and speech with increased vocal effort in the presence of various masking noise conditions, and compare these to well-known speech intelligibility-enhancing algorithms. The evaluations show that the proposed method can improve the intelligibility of synthetic speech with little loss in speech quality.
\end{abstract}
\noindent\textbf{Index Terms}: vocal effort, speech synthesis, speech-in-noise intelligibility, spectral tilt

\section{Introduction}

The intelligibility of natural and synthetic speech decreases in adverse listening conditions \cite{Hurricane1, Hurricane2, Langner05}. To improve speech intelligibility in the presence of noise, humans adapt their voice, referred to as the Lombard effect \cite{Lombard1911, Cooke2010}, by increasing the intensity of the signal, and therefore the signal-to-noise ratio (SNR), and by modifying voice characteristics as a result of increased vocal effort \cite{Lindblom1990, Lindblom1996}. Change in the vocal effort can also be triggered by the need to communicate over a distance \cite{Traunmuller00} or a change in emotional expression \cite{Ishi10, Gobl03}. Similarly, it is desirable for speech synthesizers to adapt to adverse environments and different contexts for improved intelligibility and more natural delivery of the message. Increasing the signal intensity increases the SNR and intelligibility, but this approach alone is not ideal for either the listener (discomfort) or the output device (power consumption, mechanical limits). Therefore, a preferable approach is to change the speech characteristics to increase the intelligibility, also called near-end listening enhancement \cite{Sauert2006, Skowronski2006, Yoo2007, Zorila2012}. This is especially important for text-to-speech (TTS) where the intelligibility is generally lower than in natural speech \cite{Langner05}.

Assuming an equal power constraint, there are two main approaches that have been successfully used to improve the intelligibility of synthetic speech in noise. First, using signal processing techniques, such as spectral shaping and dynamic range compression \cite{Niederjohn76, Zorila2012}, to shift the signal power to frequencies that are most relevant for intelligibility and to amplify low-energy speech sounds, such as plosives and stops. Second, mimicking human speech production in noise, which can be achieved by training or adapting a TTS synthesizer with specifically recorded Lombard or high vocal effort speech \cite{Raitio11b, RaitioSynthesis2014, QiongHu2021}.

Both approaches have been shown to improve the intelligibility of speech in noise. The signal processing based methods have the benefit that no additional speech data are required to enhance the intelligibility, and the intelligibility gains can be substantial \cite{Hurricane1, Hurricane2}. On the other hand, modeling vocal effort in speech synthesis not only increases the intelligibility of speech in noise, but also makes the synthetic voice more natural and more suitable for the listening condition \cite{Raitio11b, RaitioSynthesis2014}. However, specifically recorded Lombard or high vocal effort speech data is usually required for training the TTS system.

To overcome the limitations of requiring specifically recorded and labeled Lombard or high vocal effort speech, we propose a method that automatically estimates the vocal effort level from normal speech data using spectral tilt \cite{Campbell2003vq, Sluijter1996SpectralBA}. The natural variation of vocal effort is then modeled and controlled at synthesis time. Moreover, the model is able to extrapolate vocal effort level from the observed neutrally spoken speech data to reproduce speech with very low (soft) or high (loud) vocal effort. Also, the vocal effort modification at synthesis time has minimal effect on other prosodic factors, such as duration and pitch. Finally, the proposed method provides additional benefits, such as overall control of the prosodic space independently from other prosodic factors \cite{raitio2020prosody, raitio2022prosody}, and higher naturalness in comparison to signal processing based methods\footnote{Speech samples can be found at \href{https://apple.github.io/vocal-effort-modeling-tts-intelligibility/}{\nolinkurl{https://apple.github.io/vocal-effort-modeling-tts-intelligibility/}}.}.


\section{Background}
\label{sec:background}

The ability of speakers to adapt the speech generation process to the environment has been studied thoroughly in the past. We summarize some of the main findings, relevant to this work, in Sec.~\ref{ssec:speechmods}. The results of such studies have contributed to developing various techniques for artificially enhancing the intelligibility of natural and synthetic speech signals. To provide the context for evaluating vocal effort adjustment for intelligibility enhancement, we summarize the relevant methods in Sec.~\ref{ssec:intelliboost}.

\subsection{Speech modification for enhanced intelligibility}
\label{ssec:speechmods}

Changes in vocal effort induced by background noise are referred to as the Lombard effect \cite{Lombard1911, Junqua93, Cooke2010}. Communication over a distance \cite{Traunmuller00}, as well as emotional expression \cite{Ishi10, Gobl03} also modulate vocal effort as a means of efficient message delivery. The associated effects on naturally-produced speech have been studied, e.g., in \cite{Rostolland82, Summers88, Junqua93, Traunmuller00}.

Key speech characteristics affected in the process of modulating vocal effort include pitch, duration, intensity and spectral tilt \cite{Lombard1911}. Isolating each one of these in turn and probing its impact on intelligibility in background noise has shown that pitch and duration do not have a significant contribution \cite{Godoy2014, Hurricane2}. Such changes are explained in view of physiological constraints in the voice production mechanism. Intensity and spectral tilt, on the other hand, have a direct and significant impact on human ability to understand speech in noise.

In the case of intensity, intelligibility increases as a result of an overall increase in SNR. Reduction of spectral tilt, on the other hand, boosts the mid and high frequencies at the expense of the low frequencies. This provides an increased amount of intelligibility cues to the listener due to: 1) typical low-pass characteristics of environmental noise, and 2) the higher importance of the mid relative to the low frequency range of the speech signal \cite{OShaughnessy2000}. Fixing the average intensity and exploring time-variant as well as time-invariant frequency-dependent speech modifications has produced a numbers of methods for intelligibility enhancement. Key methods, from the perspective of TTS, are summarized in the following section.

\subsection{Methods for intelligibility enhancement of TTS signals}
\label{ssec:intelliboost}


Intelligibility enhancement of TTS can be performed at synthesis time, e.g., by parametrizing and adjusting properties of the speech generation process as a function of environment, which is the approach explored in this work. Alternatively, it can be achieved by post-processing a pre-synthesized signal. While any existing method for intelligibility enhancement can be applied in the latter case, the quality of synthetic speech must be sufficiently high to prevent signal quality degradation \cite{Botinhao2014csl}. State-of-the-art neural TTS satisfies the requirement for quality and is suitable for combination with post-processing methods.

One of the more successful post-processing methods combines dynamic range compression (DRC) with high-pass filtering \cite{Niederjohn76}. More recently, spectral shaping and dynamic range compression (SSDRC) similarly exploits the combination of a linear time-invariant (LTI) filter and DRC \cite{Zorila2012}. A multi-band parametric time-varying DRC model recovering the spectral dynamics of the signal is presented in \cite{Petkov2015}. A multi-band time-invariant DRC approach tuned to mitigate the combined effect of noise and reverberation is considered in \cite{Chermaz2020}.

Large-scale evaluations through headphones indicate that DRC-based techniques outperform alternative methodologies in terms of intelligibility gain \cite{Cooke2013sc, Hurricane2}. On the other hand, LTI and linear time variant (LTV) systems offer the benefit of lower signal processing distortion and robustness to noise in the input signal. Thus, where DRC would decrease SNR in the input signal, LTI and LTV would preserve it. Among others, some recently-proposed techniques are found in \cite{Chanda2007, Sauert2010, Taal2013}.

Intelligibility enhancement at synthesis time in the parametric space of a neural TTS system is a novel paradigm and differs from prior art due to its capability to produce both linear (filtering) and non-linear (DRC) signal modifications by non-linear transformations (neural networks). Existing work in this domain relies on reference training data, which is either created with a pre-existing signal-processing approach \cite{Paul2020}, or taken from a categorical speaking style such as clear or Lombard speech \cite{Raitio11b, RaitioSynthesis2014, GangLi2020, QiongHu2021}. The effectiveness of such methods is constrained by the quality and amount of the training data and the capability of the neural network model for adaptation. To our knowledge, this work is the first example of neural TTS providing both: 1) the capability for parametric adaptation of TTS synthesis to environmental conditions, and 2) the possibility of adaptation without dedicated and labeled training data.

\section{Modeling of vocal effort in neural TTS}
\label{sec:modeling}

In this section, we describe the measurement of vocal effort using spectral tilt, and how that can be used in the modeling of vocal effort in neural TTS.

\subsection{Measuring vocal effort using spectral tilt}

To model vocal effort without changing other prosodic aspects of speech, we extract and model the following set of prosodic features, averaged per utterance: pitch, pitch range, phone duration, speech energy, and spectral tilt. These features are easy to calculate and are robust against background noise or other recordings conditions. These features are also disentangled so that they can be varied independently \cite{raitio2020prosody, raitio2022prosody}.

The extraction of the acoustic features is described in \cite{raitio2022prosody}, however, we will describe the extraction of spectral tilt here in detail. First, we resample the speech signal to 16 kHz and use a high-pass filter with a cut-off frequency of 70~Hz to remove any low-frequency fluctuations. Then we calculate the frame-wise spectral tilt of voiced speech using the predictor coefficient of a first order all-pole filter. The spectral tilt values are averaged per utterance and then normalized to [$-$1,1] by first calculating the median ($M$) and the standard deviation ($\sigma$) and then projecting the data in the range [$M$$-$3$\sigma$, $M$$+$3$\sigma$] into [$-$1,1]. Finally, we clip values $|x|>1$ so that all data is in the range [$-$1,1].


\subsection{Neural TTS architecture}

A typical neural text-to-speech (TTS) system consists of two components: a neural front-end \cite{shen2017natural, ping2018deep, ren2020fastspeech2} that maps from phoneme input to Mel-spectrograms, and a neural back-end \cite{oord2016wavenet, kalchbrenner2018efficient} that maps from the Mel-spectrogram into a sequence of speech samples. These two networks are trained with a large amount of speech data to generate high-quality speech. Since prosody and vocal effort are mainly modeled by the neural front-end, this paper will focus on the front-end architecture.

Previously, we presented hierarchical prosody modeling and control using Tacotron 2 \cite{raitio2020prosody} and non-autoregressive parallel TTS \cite{raitio2022prosody}, both including spectral tilt modeling. In this work we focus specifically on modeling vocal effort using spectral tilt and its effect on speech intelligibility in noise.

\begin{figure}[tb]
  \centering
  \includegraphics[width=1.0\linewidth]{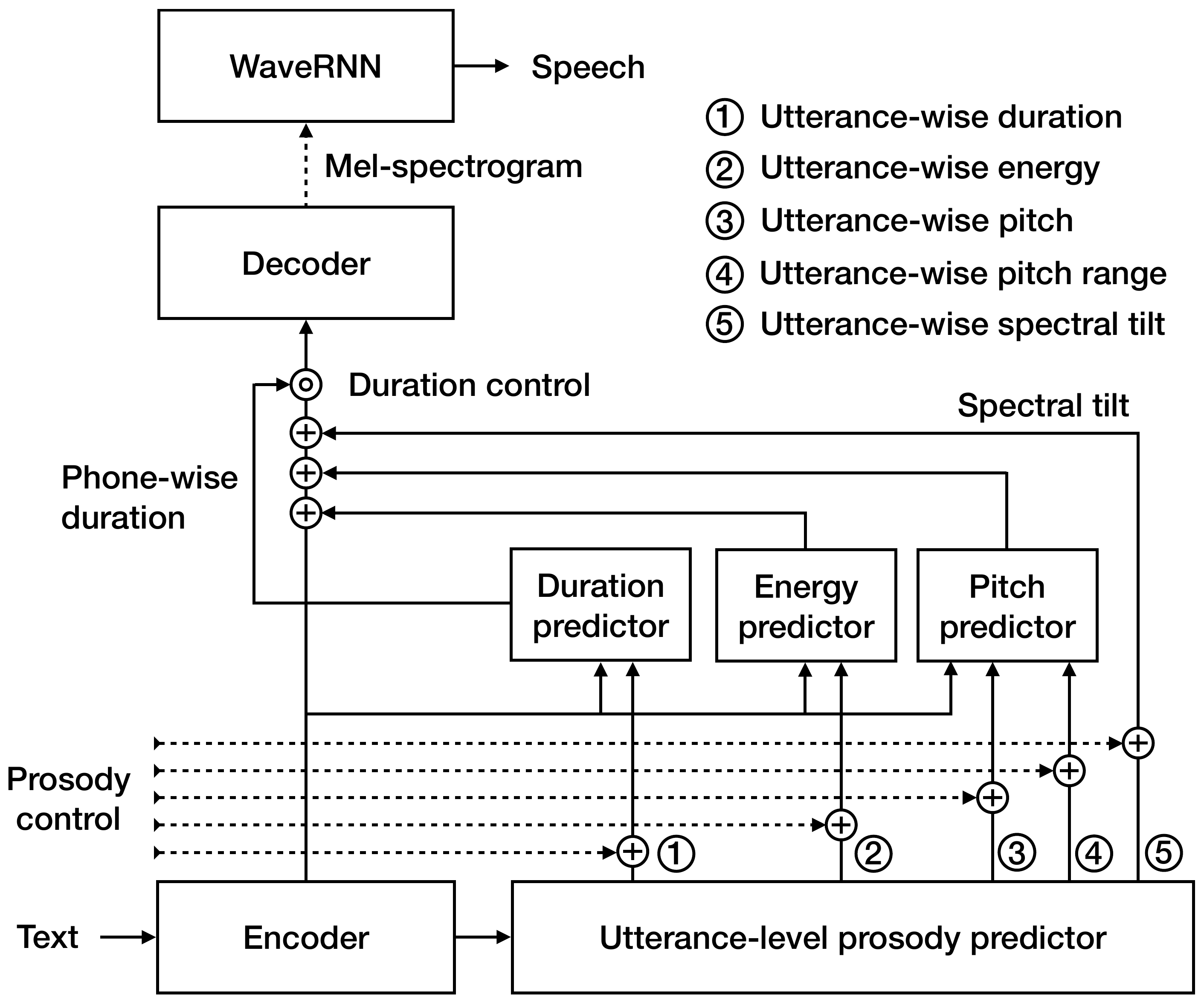}
  \vspace{-5mm}
  \caption{Proposed TTS architecture with vocal effort control.}
  \vspace{-5mm}
  \label{fig:architecture_proposed}
\end{figure}

The proposed front-end architecture, shown in Fig.~\ref{fig:architecture_proposed}, is similar to FastSpeech 2 \cite{ren2020fastspeech2}, but we extend the model with a new variance adaptor at the utterance level \cite{raitio2022prosody}. We predict the utterance-wise prosodic features from the encoder outputs, namely, pitch, pitch range, duration, energy, and spectral tilt. The architecture of the new utterance-level variance adaptor is identical to the phone-level adaptors except that we predict all five features using the same module. We use teacher-forcing of both utterance-wise and phone-wise prosodic features using ground-truth targets to efficiently train the network. At synthesis time, vocal effort control is achieved by adding bias to the spectral tilt predictions while keeping the other prosodic features unchanged to minimize undesired changes in prosody. Factorizing the prosody and only changing intended prosodic aspects is crucial in order to achieve desired changes with high quality. We observed overall lower synthesis quality with models that were conditioned only with spectral tilt due to the high correlation between speaking style and spectral tilt.

\section{Experiments}
\label{sec:experiments}

\subsection{Data}
\label{sec:data}

We use data from two American English speakers, a high-pitched speaker with 36-hour dataset (Voice 1), and a low-pitched speaker with 23-hour dataset (Voice 2). All speech data were sampled at 24 kHz. For evaluating the intelligibility and quality of synthetic speech, we used the 720 phonetically balanced Harvard sentences \cite{Harvard} as the source text for synthesis.

\subsection{Model training}
\label{sec:models}

We trained the proposed vocal effort control models for both voices. We use phone-wise duration, pitch, and energy as the fine-grained features, and utterance-wise pitch, pitch range, phone duration, speech energy, and spectral tilt as the higher-level prosodic features. 80-dimensional Mel-spectrograms are computed from pre-emphasized speech using STFT with 25~ms frame length and 10~ms shift. The encoder has 4 feed-forward Transformer layers each with a self-attention layer having 2 attention heads and 256 hidden units, and two 1-D convolution layers each having a kernel size of 9 and 1024 filters. The decoder has 2 dilated convolution blocks with six 1-D convolution layers with dilation rates of 1, 2, 4, 8, 16 and 32, respectively, kernel size of 3, and 256 filters. The feature predictors have two 1-D convolution layers each having a kernel size of 3 and 256 filters. The dropout rate of 0.2 and $\epsilon$ for layer normalization of $10^{-6}$ were used. We trained the models for 300k steps using 16 GPUs and a batch size of 512. We use WaveRNN \cite{kalchbrenner2018efficient, appleneuraltts2021asru} to generate speech from the Mel-spectrograms, trained separately for each speaker. The [$M$$-$3$\sigma$, $M$$+$3$\sigma$] spectral tilt values for Voice 1 and 2 are [$-$0.984, $-$0.926] and [$-$0.990, $-$0.931], respectively. These are mapped to [$-$1,1], where a positive value represents flatter spectral tilt and thus higher vocal effort.

\subsection{Systems}
\label{sec:systems}

We use the following systems in our evaluation:

\vspace{-1mm}
\begin{itemize}
\setlength\itemsep{-0.1mm}
\item[1.] {\bf Baseline:} Synthetic speech without modification \cite{raitio2022prosody}.
\item[2.] {\bf Proposed:} Synthetic speech with increased vocal effort.
\item[3.] {\bf SS:} Baseline processed with spectral shaping (SS) \cite{Zorila2012}.
\item[4.] {\bf SSDRC:} Baseline processed with spectral shaping and dynamic range compression (SSDRC) \cite{Zorila2012}.
\end{itemize}
\vspace{-1mm}
The baseline represents high-quality neural TTS \cite{raitio2022prosody} without any adaptation to listening conditions. The proposed system uses an increased vocal effort of +3, which means that the aimed vocal effort is $9\sigma$ higher than the median of the data. To compare the effectiveness of the proposed method to traditional approaches, we integrated two post-processing intelligibility enhancement modules to the baseline system, SS and SSDRC, which have been shown to provide state-of-the-art results \cite{Hurricane1, Hurricane2}.

The SS algorithm comprises three-stage filtering of the input speech in the short-time Fourier transform (STFT) domain. Two of the filtering stages perform phoneme-adaptive spectral shaping based on the probability of voicing, while the final filter is a fixed spectral shaper. On the adaptive spectral shaping, the local maxima (reflective of formants) are sharpened by a spectral sharpening filter followed by a high-frequency booster. Both of the filters update their coefficients based on the voicing probability of each frame. A non-adaptive pre-emphasis filter then modifies the spectra by enhancing the frequency components between 1000~Hz and 4000~Hz by a factor of 12~dB, while reducing the energy for the frequencies below 500~Hz by 6~dB/octave. Inverse Fourier transform and overlap-add are applied to get the final spectral-shaped waveform.

The DRC of the SSDRC is a time-domain operation where the objective is to reduce the envelope variation of speech. This is done by modifying the speech waveform in each frame, adaptive to the temporal envelopes. DRC is also a two-step process. In the first stage, the time-domain envelope is dynamically compressed with recursive smoothing. The smoothed envelope is projected onto the input-output envelope characteristic (IOEC) curve which gives the DRC gain. Finally, the spectral-shaped waveform from the SS module is multiplied by the estimated gains in the DRC to generate the SSDRC waveform.

\subsection{Objective measures}
\label{sec:objective}

\begin{figure}[tb]
  \centering
  \includegraphics[width=1.0\linewidth]{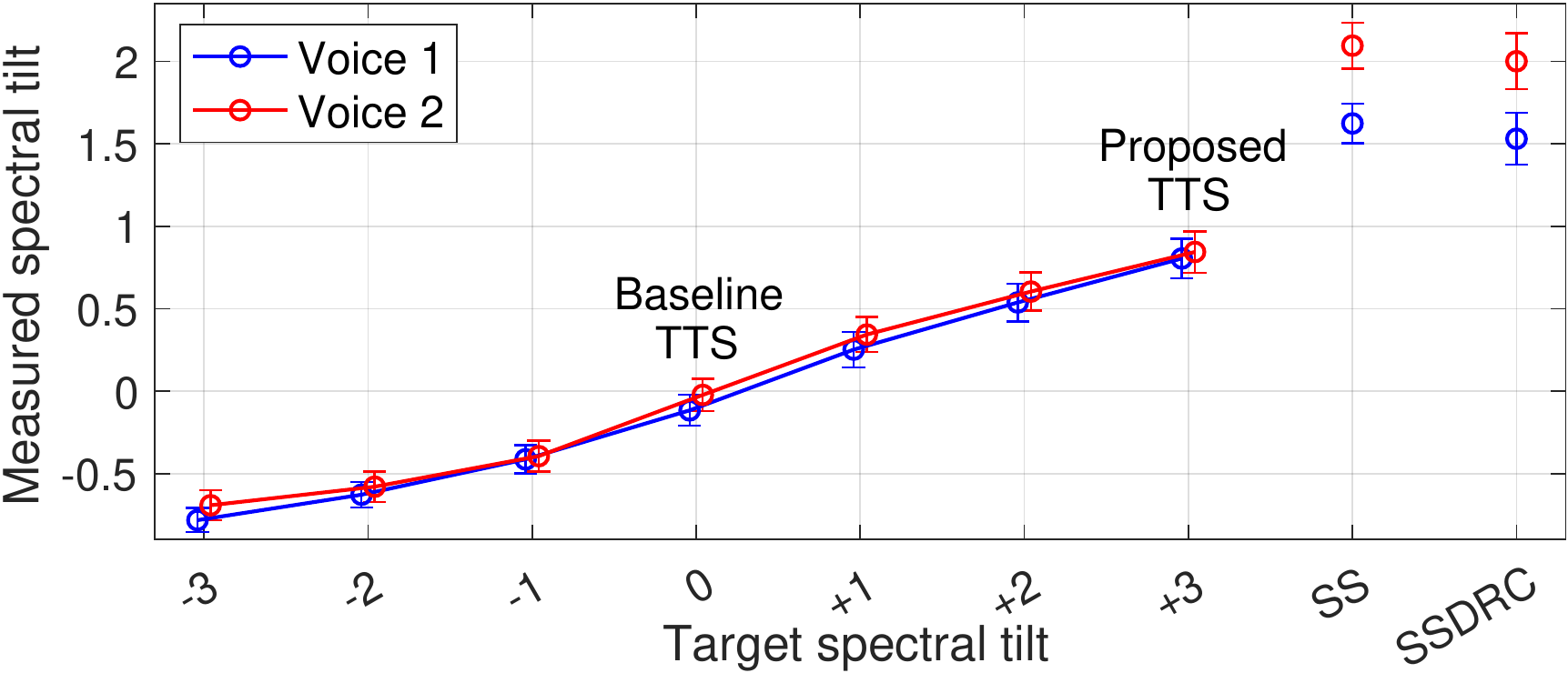}
  \vspace{-6mm}
  \caption{Means and 95\% confidence intervals of normalized spectral tilt (vocal effort) measured for all systems, including various target spectral tilt values for the proposed TTS system.}
  \vspace{-5mm}
  \label{fig:tilt}
\end{figure}

\begin{figure}[tb]
  \centering
  \includegraphics[width=1.0\linewidth]{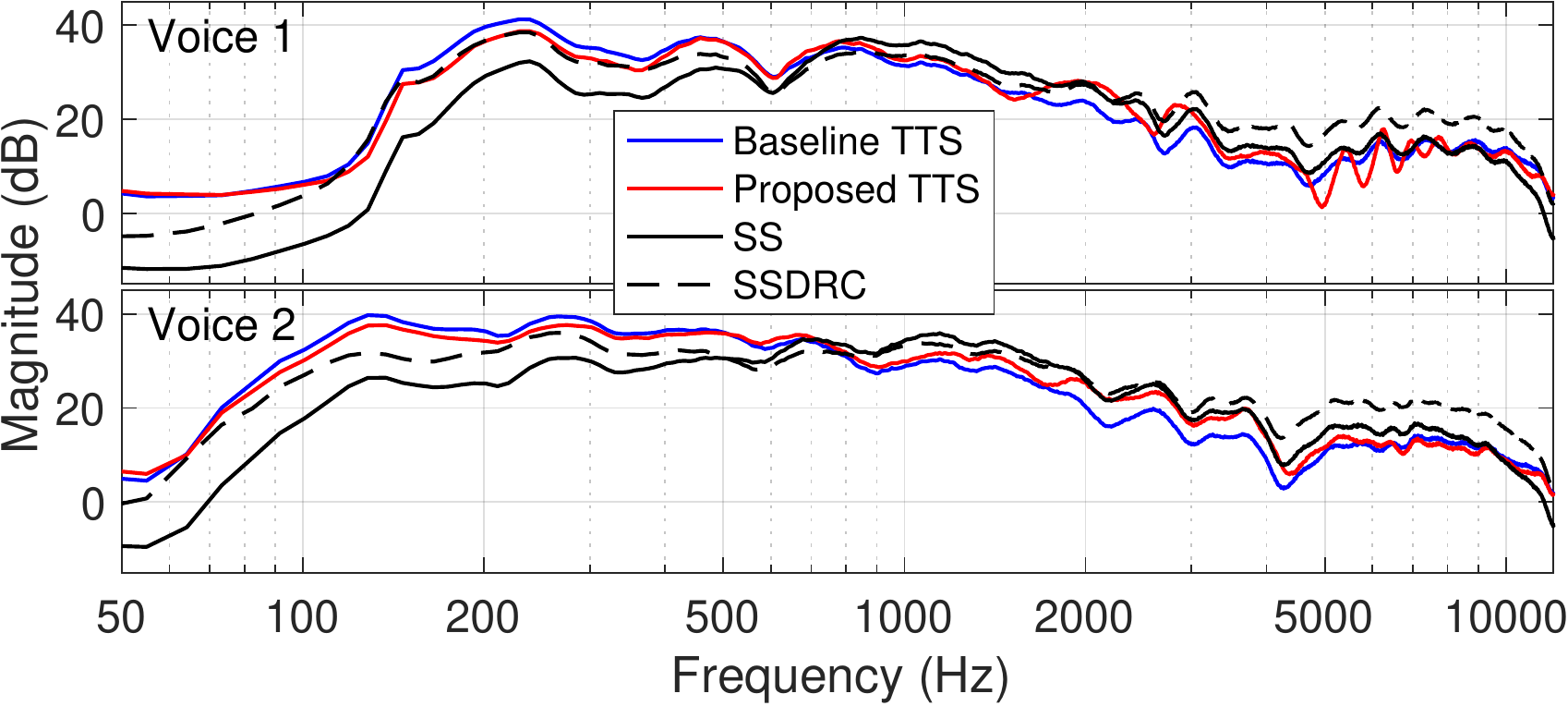}
  \vspace{-7mm}
  \caption{Long-term average spectra over all speech samples.}
  \vspace{1mm}
  \label{fig:spectra}
\end{figure}

We used the 720 Harvard utterances \cite{Harvard} to synthesize speech from each system and voice. In addition, to demonstrate the change in the vocal effort in the proposed system, we generated synthetic speech with various target spectral tilt values from -3 to +3, and then measured the output spectral tilt. The results, illustrated in Fig.~\ref{fig:tilt}, show that the (normalized) spectral tilt, and therefore the vocal effort, increases as the target vocal effort is increased. However, the SS and SSDRC systems have higher (flatter) spectral tilt in comparison to the proposed TTS system with adjusted spectral tilt of +3. We also calculated the long-term average spectra (LTAS) over all the speech samples for each system and voice. The LTAS, shown in Fig.~\ref{fig:spectra}, illustrate that all methods show a shift of energy from low to mid and high frequencies in comparison to the baseline.

\subsection{Speech intelligibility in noise}

We generated speech samples for each system and each speaker using the 720 the Harvard utterances \cite{Harvard}. We used two types of masking noises in the test: a fluctuating masker using competing speaker (CS) and a stationary masker using speech shaped noise (SSN). The CS masker was recorded read speech with alternate voice (Voice 2 for evaluating Voice 1, and vice versa) with SNRs of -7, -14, and -21 dB. The SSN masker was constructed by measuring the LTAS of the alternate voice with linear prediction and then filtering white noise with the resulting spectrum. The SNRs used were 1, -4, and -9 dB. The active speech level of the target samples were normalized according to ITU-T P.56 standard \cite{ITU93} before adding the masking noise according to the SNRs. The target speech content was centrally embedded in the noise with 0.5 seconds of lead and lag time. Overall, the test design was similar to \cite{Hurricane1}.

\begin{figure}[tb]
  \centering
  \vspace{-2.7mm}
  \includegraphics[width=1.0\linewidth]{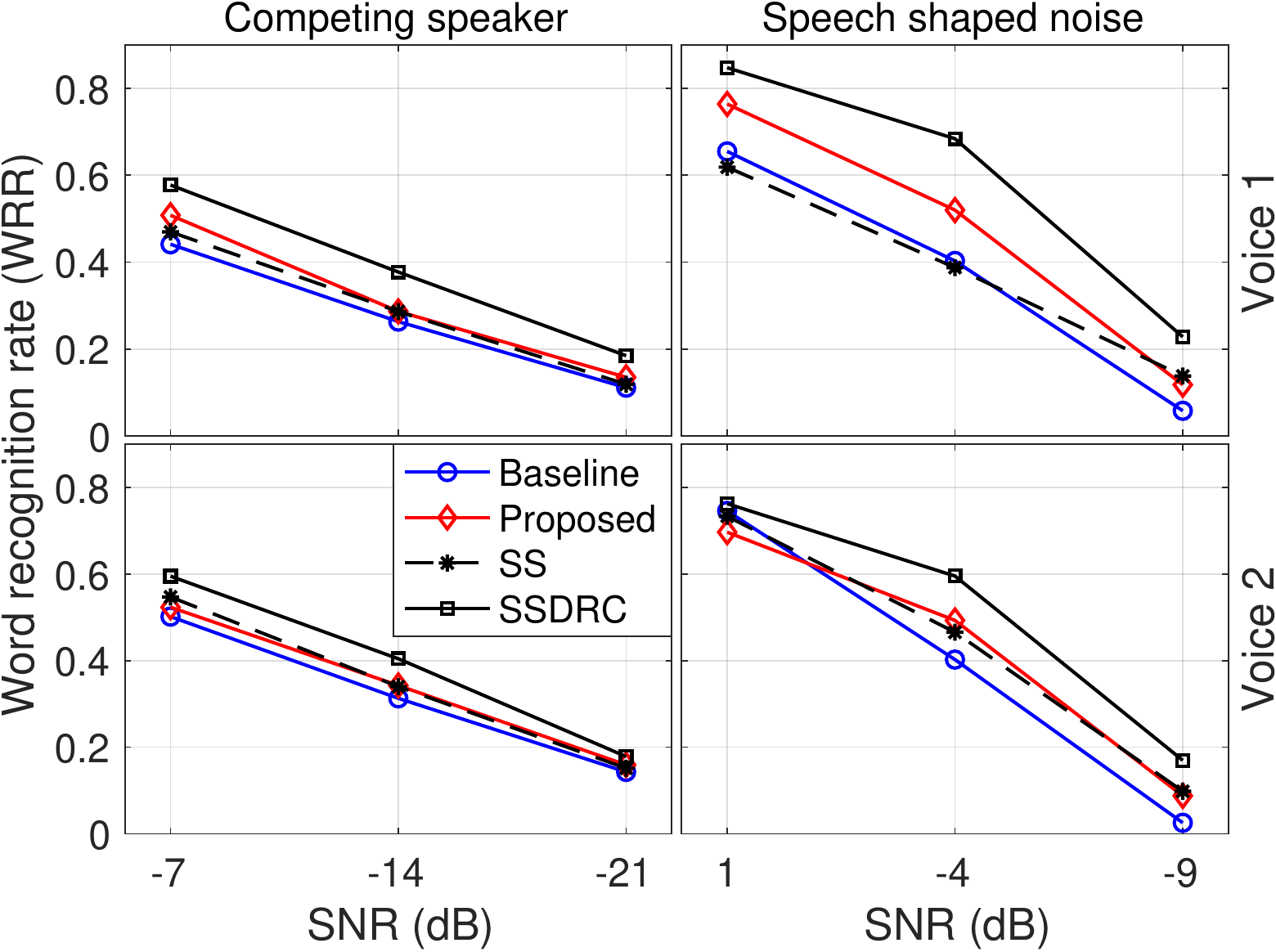}
  \vspace{-6mm}
  \caption{Word recognition rates (WRR) for all conditions.}
  \vspace{-5mm}
  \label{fig:wrr}
\end{figure}

We organized a large intelligibility test where test subjects used a graphical UI and were asked to listen to a speech sample in noise only once, and then type what they heard. Each listener was restricted to evaluate only a maximum of 720 utterances so that they never heard the same target utterance twice. The total number of samples in the test was 34,560, consisting of 2 voices $\times$ 4 systems $\times$ 2 noise types $\times$ 3 SNRs $\times$ 720 utterances. We added 10 reference speech samples without noise for each listener to evaluate their performance. The listeners had to achieve 80\% word recognition rate (WRR) on the reference samples and more than 10\% WRR on the test samples in order to qualify.

A total of 29,520 ratings were given by 41 qualified listeners using earbuds (47\%), headphones (38\%), or loudspeakers (16\%). The WRR was measured as the proportion of content words that were recognized correctly and in the right order. The intelligibility test results are shown in Fig.~\ref{fig:wrr}.

Linear mixed effects models with {\it system} as the fixed effect and {\it voice}, {\it listener}, {\it utterance} and {\it audio device} as random intercepts were fit to the data using the lme4 package \cite{lme4}. Comparing models with and without {\it system}, a likelihood ratio test ($\chi^2$ = 1105.2, p$<$.0001) and a drop in Akaike information criterion showed that {\it system} significantly predicted WRR. Dunnett's multiple comparison test from the emmeans package \cite{emmeans} shows that the proposed system had significantly better intelligibility than baseline and SS on average (p$<$.0001), while SSDRC had significantly better intelligibility (p$<$.0001) on average than other systems. As expected, the audio device had a clear effect on intelligibility, with the following average WRRs: earbuds: 37.9$\pm$0.6~\%, headphones 42.5$\pm$0.7~\%, speakers: 32.8$\pm$1.0~\%.


\subsection{Quality without noise}

To measure the effect of the intelligibility enhancement on the overall quality of speech without noise, we performed a mean opinion score (MOS) test with each of the 4 systems and 2 speakers using the 720 Harvard sentences. A 5-point MOS test was performed by 68 individual American English native speakers (51 female, 17 male, 18-69 years old) using either headphones (33) or loudspeakers (35). Listeners provided 5 ratings per utterance, resulting in a total of 28,800 responses.

\begin{figure}[tb]
  \centering
  \vspace{-1mm}
  \includegraphics[width=1.0\linewidth]{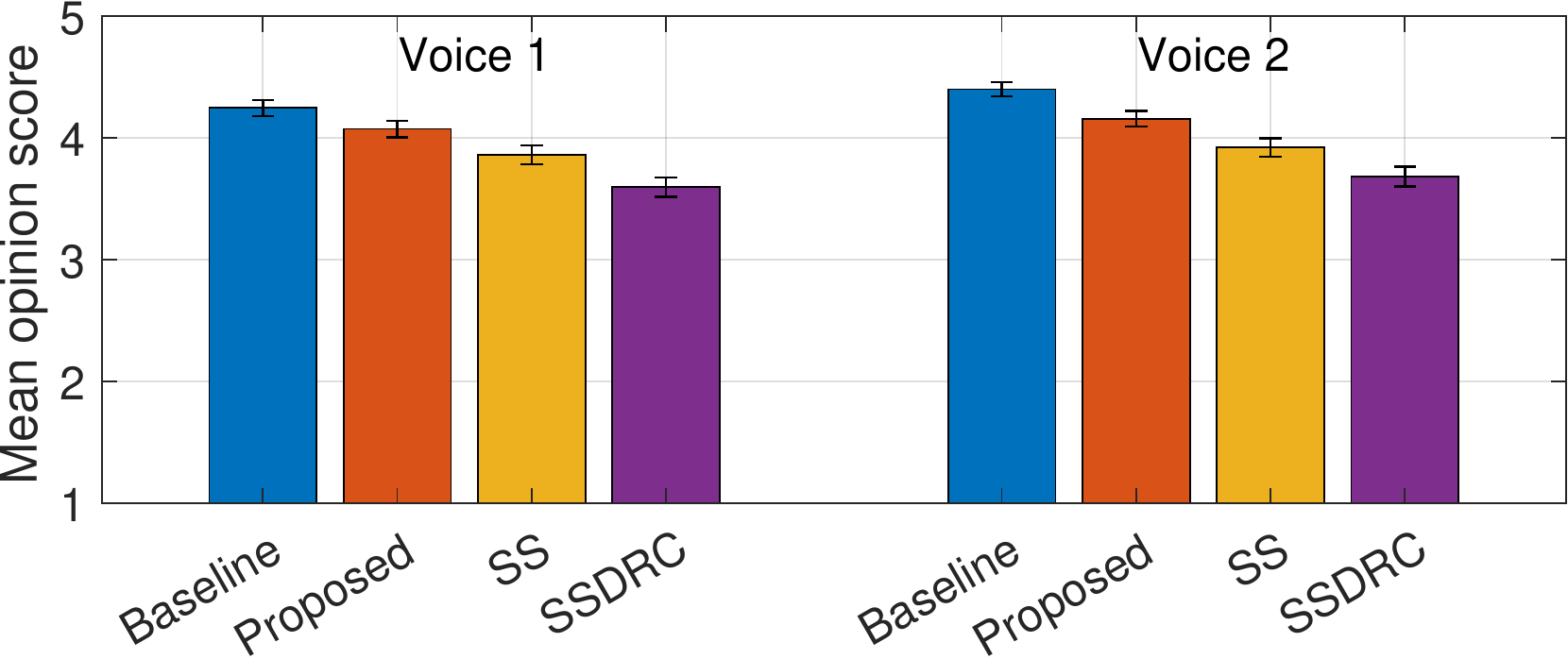}
  \vspace{-4mm}
  \caption{Means and 95\% confidence intervals of MOS.}
  \vspace{-3.5mm}
  \label{fig:mos}
\end{figure}

The results are shown in Fig.~\ref{fig:mos}. Linear mixed effects models with {\it system} and {\it voice} as fixed effects and {\it listener} and {\it utterance} as random intercepts were fit to the data. Likelihood ratio tests showed no significant interaction between {\it system} and {\it voice} (p=.57), and significant effects of {\it system} (p$<$.0001). Pairwise comparisons using the Tukey adjustment method showed significant differences between all systems (p$<$.0001). In the absence of noise, synthetic speech without any modification is preferred by listeners, which we would expect, but the proposed method was significantly preferred above other intelligibility enhancement methods (across both voices).

\section{Discussion and conclusions}
\label{sec:conclusions}


The focus of our experiment was on the average effect of intelligibility, and we do observe a consistent directional improvement for the proposed method across SNRs for both voices. Nonetheless, the data suggest that the strength of the effect may be contextual. In particular, the difference of most systems from baseline seems to generally be higher for Voice 1 than Voice 2. It may be that intelligibility enhancement methods work better with some voices than others. Our experiment was designed with sufficient power to detect main effects only, as that is the focus of this paper, and thus any exploration of the effect of context or voice is more speculative and worthy of exploration in future work. Also, developing a framework for automatically determining optimal vocal effort level with respect to masking noise type and level is left for future work.



We proposed a neural TTS system that models the natural vocal effort variation in the training data without specifically designed or labeled database. We use spectral tilt to measure vocal effort, and condition our models with spectral tilt among other prosodic features to factorize the prosodic space. By altering the input spectral tilt at inference time, we can successfully control the vocal effort of synthetic speech. Our experiments show that the proposed system achieved significantly higher word recognition rate than the baseline, while in the absence of noise, the proposed method achieves higher quality than other intelligibility enhancement methods.



\newpage
\bibliographystyle{IEEEtran}
\bibliography{references}

\end{document}